# Bohmian Mechanics vs. Standard Quantum Mechanics: a Difference in Experimental Predictions.

Artur Szczepański
Retired from the Institute of Fundamental Technological Research, Warszawa, Poland

**Abstract**
Standard Quantum Mechanics (QM) predicts an anti-intuitive phenomenon here referred to as "quantum autoscattering", which is excluded by Bohmian Mechanics. The scheme of a *gedanken* experiment testing the QM prediction is briefly discussed.

The physics of the beam splitter action upon a single quanton is different in Bohmian Mechanics (BM) and in standard Quantum Mechanics (QM). According to the QM formalism the wave packet of a single quanton is split in parts consisting of the one particle and the vacuum state superposition. The BM picture is different since the particle aspect of the quanton is not split: it is either reflected or transmitted (For a recent overview of BM see, e.g. [1]). Therefore it is quite natural to look for discrimination between BM and QM in experiments with split quantum objects. Interferometry of single neutrons subject to spin-flipping inside the interferometer has been thought to supply the means [2]. However, the experimental results have not been interpreted as providing clear-cut conclusions.

An interference pattern forms when the split parts of a single quanton recombine. In QM this is not the only nontrivial result of the recombination. The overlapping of both the partial wave packets from the reflected and the transmitted beam generates with a nonvanishing probability scattering states of the one-particle contributions, referred hereafter to as autoscattering states. By contrast, autoscattering states are obviously excluded in BM.

Here I consider a *gedanken* experiment aimed at observing autoscattering states of split electron wave packets. Its scheme is shown in Fig. 1 below.

A low-intensity beam of electrons is split by a Stern-Gerlach magnet into a spin-up and a spin-down partial beam. The partial beams are deflected without affecting the spin by a focusing system to recombine. Energy-momentum conservation allow for only one autoscattering channel predicted by QM: the exchange scattering channel with scattered packets directed along the non-scattered outgoing paths. The spin-orientation difference allow for selecting and detecting the autoscattered electrons. On the other hand, according to BM no autoscattered electrons should be observed: in a real experiment the count rate of the detector $D2$ and $D4$ (see, Fig. 1) should not exceed the noise level.

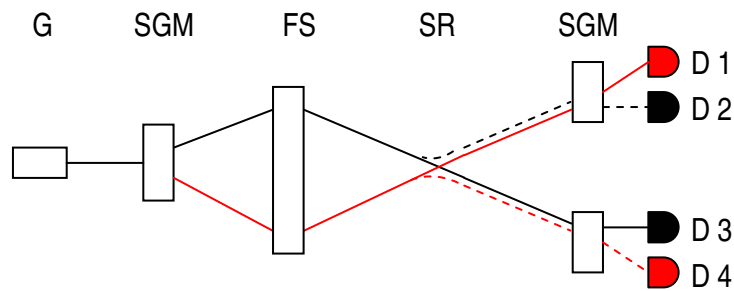

Fig. 1. Schematic of the experiment. The wave packet of a single electron emitted by the gun, G, is split by the Stern-Gerlach magnet, SGM. The spin up partial packet travels along the, say, upper (black) path, while the spin down partial packet travels along the lower (red) path. Both the packets are directed by a focusing system, FS, into the recombination (scattering) region, SR. Outgoing wave packets are spin-analyzed and detected. Broken lines indicate the paths of autoscattered wave packets according to the prediction of standard quantum mechanics.

Conceptually simple experimental ideas are usually technically hard to realize. Note that the proposed scheme while feasible would require stringent stability and alignment conditions to achieve a sufficient statistics of partial-wave-packets-good-overlapping events in the scattering region.

___________
Correspondence address: artallszczep@gmail.com